\documentclass[apj]{emulateapj}

\newcommand{\etal}{{et al.~}}
\newcommand{\msunh}{\>h^{-1}\rm M_\odot}
\newcommand{\Msun}{\>{\rm M_\odot}}

\newcommand{\mpch}{\>h^{-1}{\rm {Mpc}}}
\newcommand{\kms}{\>{\rm km}\,{\rm s}^{-1}}
\newcommand{\calC}{{\cal C}}
\newcommand{\rmag}{\>^{0.1}{\rm M}_r-5\log h}

\newcommand{\Mh}{M_{\rm h}}

\newcommand{\Mstar}{M_{\ast}}
\newcommand{\calPc}{{\cal P_{\rm c}}}

\newcommand{\kmsmpc}{\>{\rm km}\,{\rm s}^{-1}\,{\rm Mpc}^{-1}}
\newcommand{\dd}{\mathrm{d}}

\def\gtsima{$\; \buildrel > \over \sim \;$}
\def\ltsima{$\; \buildrel < \over \sim \;$}
\def\gta{\lower.7ex\hbox{\gtsima}}
\def\lta{\lower.7ex\hbox{\ltsima}}

\shorttitle{The stellar mass components} \shortauthors{Liu et
al.}

\begin{document}


\title{The   Stellar   Mass   Components  of   Galaxies:\\
  Comparing Semi-Analytical Models with Observation}

\author{Lei  Liu\altaffilmark{1,5}, Xiaohu  Yang\altaffilmark{1}, H.J.
  Mo \altaffilmark{2}, Frank  C. van den Bosch\altaffilmark{3}, Volker
  Springel\altaffilmark{4}}

\altaffiltext{1}{Key   Laboratory  for   Research   in  Galaxies   and
  Cosmology, Shanghai  Astronomical Observatory; the  Partner Group of
  MPA;   Nandan    Road   80,   Shanghai    200030,   China;   E-mail:
  liulei@shao.ac.cn}
\altaffiltext{2}{Department of Astronomy, University of Massachusetts,
  Amherst MA 01003-9305}
\altaffiltext{3} {Department  of Physics and  Astronomy, University of
  Utah, 115 South 1400 East, Salt Lake City, UT 84112-0830}
\altaffiltext{4}      {Max-Planck-Institut      f\"ur     Astrophysik,
  Karl-Schwarzschild-Strasse 1, 85748 Garching, Germany}
\altaffiltext{5}{Graduate School  of the Chinese  Academy of Sciences,
  19A, Yuquan Road, Beijing, China}


\begin{abstract}
  We  compare the  stellar masses  of central  and  satellite galaxies
  predicted   by  three   independent   semi-analytical  models   with
  observational results  obtained from a large  galaxy group catalogue
  constructed  from the Sloan  Digital Sky  Survey. In  particular, we
  compare the  stellar mass functions of centrals  and satellites, the
  relation  between  total  stellar   mass  and  halo  mass,  and  the
  conditional   stellar  mass  functions,   $\Phi(\Mstar|\Mh)$,  which
  specify the average number of galaxies of stellar mass $\Mstar$ that
  reside in  a halo  of mass $\Mh$.   The semi-analytical  models only
  predict  the correct  stellar masses  of central  galaxies  within a
  limited  mass range  and  all  models fail  to  reproduce the  sharp
  decline of  stellar mass with  decreasing halo mass observed  at the
  low mass  end.  In addition,  all models over-predict the  number of
  satellite galaxies by roughly a factor of two. The predicted stellar
  mass in satellite galaxies can be made to match the data by assuming
  that  a  significant  fraction  of satellite  galaxies  are  tidally
  stripped and disrupted, giving rise to a population of intra-cluster
  stars  in their host  halos.  However,  the amount  of intra-cluster
  stars  thus predicted is  too large  compared to  observation.  This
  suggests  that current  galaxy formation  models still  have serious
  problems in modeling star formation in low-mass halos.
\end{abstract}




\section{Introduction}
\label{sec_intro}

Semi-analytical models (hereafter SAMs) are a powerful method to study
the  formation and  evolution of  galaxies in  a CDM  cosmogony (e.g.,
White  \& Frenk  1991; Lacey  \etal 1991;  Kauffmann \etal  1993; Cole
\etal 1994,  2000; Somerville \etal  1999; Benson \etal  2000).  Since
dark  matter  couples to  baryons  only  through  gravity, the  galaxy
formation process is expected to only  have a small effect on the dark
matter   distribution.    Tests   using  N-body   and   hydrodynamical
simulations  (with and  without star  formation) have  shown  that the
general properties of  the large scale structure, such  as the overall
structure and distribution of dark matter halos, are not significantly
affected by  gas physics  (e.g.  Lin et  al.  2006).  It  is therefore
possible to  separate the modeling  of galaxy formation  and evolution
into two steps:  (i) modeling the formation and  evolution of the halo
population  using  either   N-body  simulations  or  (semi)-analytical
methods (i.e., extended Press-Schechter theory), and (ii) modeling how
galaxies form and  evolve within individual dark matter  halos using a
semi-analytical  approach.   In  the  second  step,  one  incorporates
various physical  processes, such as  gas cooling, star  formation and
feedback, to predict the properties of the galaxy population.

The  advantage  of  this  semi-analytical approach  compared  to,  for
example,  full   hydrodynamical  simulations  is  that   it  allows  a
relatively fast  and flexible exploration of a  large parameter space.
Typically,  the  free parameters  that  describe  the efficiencies  of
cooling, star  formation and feedback  are tuned to  reproduce certain
(global) observational  constraints, such as  the luminosity function,
the  Tully-Fisher  relation and  the  color-magnitude relation,  among
others. In the past decade this technique has been used extensively to
constrain  the various  physical processes  that  play a  role in  the
formation  and evolution  of  galaxies, and  to  make predictions  for
future  observations (e.g.,  Kauffmann \etal  1999; Helly  \etal 2003;
Kang \etal 2005;  Bower \etal 2006; Croton \etal  2006; Cattaneo \etal
2006;  De Lucia  \& Blaizot  2007; Monaco,  Fontanot \&  Taffoni 2007;
Henriques, Bertone \& Thomas  2008; Somerville \etal 2008; Neistein \&
Weinmann 2009).

So far,  however, in most  of the SAMs,  only global properties of  the galaxy
population have been used extensively and consistently as constraints on their
free parameters.  Although reproducing these  global properties is  clearly an
important first  step, it  lacks the power  to constrain model  assumptions in
detail.  For example,  a model may overestimate the  galaxy population in some
(e.g.   massive)   halos  while  underestimating  that  in   other  halos,  or
overestimate the  stellar mass in central galaxies  while underestimating that
in satellites, and yet match the total stellar mass function.

In  recent years,  much progress  has  been made  in constraining  the
properties of galaxies as function of halo mass. For example, numerous
authors have  used the clustering  properties of galaxies in  order to
constrain  the halo occupation  distribution (HOD),  $P(N|\Mh)$, which
specifies  the probability  that a  halo  of mass  $\Mh$ contains  $N$
galaxies (e.g., Jing, Mo \& B\"orner 1998; Scranton 2003; Zehavi \etal
2004, 2005;  Tinker \etal 2005;  Collister \& Lahav 2005;  Zheng \etal
2005;  2007; Tinker  \&  Wetzel 2009)  or  the conditional  luminosity
function (CLF), $\Phi(L|\Mh)\dd L$, which specifies the average number
of galaxies of luminosity $L\pm\dd L/2$  that reside in a halo of mass
$\Mh$ (e.g., Yang, Mo \& van den Bosch 2003; van den Bosch, Yang \& Mo
2003;  Cooray   2006;  van  den  Bosch  \etal   2007;  Cacciato  \etal
2009).  These  statistics   provide  detailed  information  about  how
galaxies of  different luminosities (or stellar  masses) are connected
to dark matter  halos of different masses, and  can therefore put more
stringent constraints on models of galaxy formation and evolution.

However, as  pointed out  by Yang \etal  (2005b), one  disadvantage of
these  HOD/CLF models  is that  the results  are not  completely model
independent, i.e. one typically has to postulate a functional form for
either $P(N|\Mh)$ or $\Phi(L|\Mh)$.   This problem can be circumvented
by using galaxy group catalogues.  If galaxy groups are defined as the
ensembles of galaxies  that reside in the same  dark matter host halo,
these  group  catalogues  yield  a  much  more  direct  probe  of  the
galaxy-dark  halo connection. With  this in  mind, Yang  \etal (2005a)
developed an  adaptive halo-based group finder that  can properly link
galaxies  according to  their common  dark matter  halos.   Yang \etal
(2007) applied this  halo-based group finder to the  Sloan Digital Sky
Survey (SDSS)  Data Release 4 (Adelman-McCarthy \etal  2006), and used
the  resulting group  catalogues to  infer the  conditional luminosity
functions  (CLF) and  the  conditional stellar  mass functions  (CSMF)
directly from the data,  separately for central and satellite galaxies
(Yang \etal  2008, 2009b; hereafter  Y09b). Especially the  ability to
split the galaxy population in centrals and satellites is an important
advantage of  using group  catalogues.  After all,  from the  point of
view of galaxy formation, central and satellite galaxies are subjected
to very different processes:  whereas central galaxies are believed to
reside  at  the  centers  of  their  dark  matter  halos,  where  they
cannibalize satellite  galaxies that have  lost their momentum  due to
dynamical friction, and  act as the recipients of  new gas via cooling
flows,  satellite  galaxies  orbit  around central  galaxies  and  are
subjected to  a number of satellite-specific processes,  such as tidal
stripping and heating,  ram-pressure stripping, galaxy harassment, and
strangulation. Consequently, central and satellite galaxies of a given
stellar mass are expected to have different properties, something that
has  recently  been confirmed  observationally  using  the SDSS  group
catalogues  of Y07  (e.g., van  den Bosch  \etal 2008;  Pasquali \etal
2009a,b; Weinmann \etal 2006a, 2009; Skibba 2009).

A  comparison of the  halo occupation  statistics obtained  from these
galaxy group  catalogues with predictions  from semi-analytical models
has already provided important  new insights into galaxy formation and
evolution.  Weinmann \etal (2006b)  and Kimm \etal (2009) compared the
color  distributions of  central and  satellite galaxies  in  halos of
different masses  to predictions from  various semi-analytical models,
and showed that the  latter dramatically over-predict the red fraction
of  satellite  galaxies.   This   problem  has  become  known  as  the
over-quenching problem, and has triggered a number of studies into the
mechanisms  that may cause  quenching of  star formation  in satellite
galaxies (e.g.,  Baldry \etal 2006; Kang  \& van den  Bosch 2008; Font
\etal 2008;  McCarthy \etal 2008;  van den Bosch \etal  2008; Fontanot
\etal 2009; Weinmann \etal  2009).  Yang \etal (2009a; hereafter Y09a)
used the CSMF  obtained from the group catalogues  to discuss the fate
of satellite  galaxies, and suggested  that a significant  fraction is
likely to  be tidally disrupted  after being accreted into  their host
halos, producing a population  of intra-cluster stars.  Pasquali \etal
(2009b) studied  the ages and  metallicities of central  and satellite
galaxies as functions  of both stellar mass and  halo mass, and showed
that  the  SAM of  Wang  \etal  (2008),  which predicts  stellar  mass
functions  and   two-point  correlation  functions   in  good  overall
agreement  with  observations,  fails   to  reproduce  the  halo  mass
dependence of  the metallicities of low mass  satellites, and predicts
that satellite galaxies have the same metallicities as centrals of the
same stellar  mass, in disagreement  with the data. In  agreement with
Y09a,  they  argue  that this  is  likely  to  reflect the  impact  of
satellite disruption, a process that has almost never been included in
semi-analytical models thus far (but see Benson \etal 2002).

In this  paper we make  use of the  various stellar mass  functions of
central and  satellite galaxies in halos of  different masses obtained
by  Y09b to  evaluate  three recent  SAMs  carried out  by Kang  \etal
(2005), Bower \etal (2006) and De Lucia \& Blaizot (2007).  This paper
is  organized as  follows.  In  \S\ref{sec_SDSS} we  outline  the main
properties of  the SDSS DR4 galaxy  group catalogues.  \S\ref{sec_SAM}
gives a brief  description of the semi-analytical models  used in this
paper, highlighting  their similarities as well  as their differences.
In  \S\ref{sec_MGRS},  we describe  the  construction  of mock  galaxy
redshift  surveys and  the  corresponding mock  group catalogues.   In
\S\ref{sec_comparing}  we compare the  galaxy stellar  mass functions,
the  relation  between total  stellar  mass  and  halo mass,  and  the
conditional  stellar mass functions  predicted by  the semi-analytical
models with the  results from the Y07 group  catalogue. We discuss the
implications of this  comparison in \S\ref{sec_missing}, and summarize
our results in \S\ref{sec_sum}.

Throughout this paper  we adopt a $\Lambda$CDM cosmology  with parameters that
are consistent with the three-year data release of the WMAP mission (hereafter
WMAP3  cosmology):  $\Omega_{\rm  m}  = 0.238$,  $\Omega_{\Lambda}  =  0.762$,
$\Omega_{\rm  b}  =  0.042$,  $n=0.951$,  $h=H_0/(100  \kmsmpc)  =  0.73$  and
$\sigma_8 = 0.75$ (Spergel \etal 2007).  Wherever necessary, we have converted
the halo masses to this particular cosmology using abundance matching based on
the halo mass functions. Note however, we did not adjust any galaxy properties
accroding to  the updated halo  masses, i.e., rerun  the SAMs. Since  SAMs are
constrained using the global properties of galaxies, we expect that the change
will be small and not impact any of our results significantly.

\section{Galaxy groups in SDSS DR4}
\label{sec_SDSS}

The  observational   data  used  here  are   galaxy  group  catalogues
constructed from the New  York University Value-Added Galaxy Catalogue
(NYU-VAGC;  Blanton \etal  2005),  which  is based  on  the SDSS  Data
Release  4  (Adelman-McCarthy \etal  2006).   From  this NYU-VAGC  Y07
selected all  galaxies in  the Main Galaxy  Sample with  an extinction
corrected apparent  magnitude brighter than $r=18$,  with redshifts in
the range  $0.01 \leq  z \leq 0.20$  and with a  redshift completeness
$\calC_z >  0.7$. This sample of  galaxies is used  to construct three
group samples:  sample I,  which only uses  the 362,356  galaxies with
measured redshifts from the SDSS,  sample II which also includes 7,091
galaxies  with   SDSS  photometry   but  with  redshifts   taken  from
alternative  surveys,  and sample  III  which  includes an  additional
38,672  galaxies that  lack a  redshift due  to  fiber-collisions, but
which  we assign  the redshift  of its  nearest neighbor  (cf.  Zehavi
\etal 2002).  The analysis presented  in this paper is mainly based on
sample  II.  Survey  edge  effects  have been  taken  into account  by
removing those groups (about 1.6\% of the total) that are too close to
one of  the edges of the  survey. The stellar mass,  $\Mstar$, of each
galaxy is  computed using the relations  between stellar mass-to-light
ratio and $^{0.0} (g-r)$ color from Bell \etal (2003),
\begin{eqnarray}\label{eq:MtoL}
\log\left[\frac{\Mstar}{h^{-2}\Msun}\right]
& = & -0.306 + 1.097 \left[^{0.0}(g-r)\right]-0.10 \nonumber\\
&   & -0.4(^{0.0}M_r - 5\log h - 4.64)\,.
\end{eqnarray}
Here  $^{0.0}(g-r)$ and $^{0.0}M_r-5\log  h$ are  the $g-r$  color and
$r$-band magnitude $K+E$-corrected  to $z=0$, respectively, the number
$4.64$ is the $r$-band magnitude of  the Sun in the AB system (Blanton
\& Roweis  2007), and  the $-0.10$ term  reflects the assumption  of a
Kroupa (2001) IMF.

Galaxies are  split into ``centrals'',  which are defined as  the most
massive   group  members  in   terms  of   their  stellar   mass,  and
``satellites'', which  are those group members that  are not centrals.
For each group  in the Y07 catalogue two estimates  of its dark matter
halo mass, $\Mh$, are available: one based on the ranking of its total
characteristic luminosity, and  the other based on the  ranking of its
total characteristic  stellar mass. Both  halo masses agree  very well
with each  other, with  an average scatter  that decreases  from $\sim
0.1$~dex  at  the low  mass  end to  $\sim  0.05$~dex  at the  massive
end.  With the  method of  Y07, halo  masses can  only be  assigned to
groups more  massive than  $\sim 10^{12} h^{-1}  \Msun$ which  have at
least  one member  with $^{0.1}M_r  - 5  \log h  \leq$ -19.5  mag. For
smaller mass halos, Yang \etal  (2008) have used the relations between
the luminosity (stellar mass) of central galaxies and the halo mass of
their groups to  extrapolate the halo mass of  single central galaxies
down to $\Mh \simeq 10^{11}  h^{-1} \Msun$. This extends the number of
galaxies with an  assigned halo mass from 295,861  in the original Y07
paper to all 369,447 galaxies in sample II.

Due to  the flux limit  of the survey,  only galaxies brighter  than a
certain magnitude can be  observed. This induces incompleteness in the
stellar masses of galaxies and the halo masses of groups.  As shown in
the Appendix of van den Bosch  \etal (2008), for the stellar masses of
galaxies, the  apparent magnitude limit  of the galaxy sample,  $m_r =
17.77$,  can be  translated to  a stellar  mass limit  as  function of
redshift $z$:
\begin{eqnarray} \label{eq:mstarlim}
\lefteqn{\log[M_{\ast,{\rm lim}}/(h^{-2}\Msun)] =} \\
& & {4.852 + 2.246 \log D_L(z) + 1.123 \log(1+z) - 1.186 z
\over  1 -  0.067  z}\,. \nonumber
\end{eqnarray}
The  galaxy  sample  is   complete  for  galaxies  with  $\Mstar  \geq
M_{\ast,{\rm  lim}}$. As shown  in Y09b,  the corresponding  halo mass
limit at $z$, $\log M_{\rm h, lim}$, is given by
\begin{equation}\label{eq:Mh_limit}
\log M_{\rm h, lim} =(z-0.085)/0.069 + 12\,.
\end{equation}
The group  catalogue is complete for  groups with $\Mh  \geq M_{\rm h,
  lim}$.  Taking these  two mass  limits into  account,  Y09b measured
various stellar mass functions  for galaxies and groups, including the
conditional stellar mass  functions $\Phi(\Mstar|\Mh)$.  In this paper
we   use  these   statistics   to  evaluate   the  three   independent
semi-analytical models described below.

\section{Semi-analytical models}
\label{sec_SAM}

The first semi-analytical model to  be considered is the one presented
in De Lucia  \& Blaizot (2007; hereafter D07),  which uses the methods
developed by  Kauffmann \& Haehnelt (2000), Springel  \etal (2001) and
De  Lucia \etal  (2006).  This  model is  a modified  version  of that
presented in Croton \etal (2006),  and includes a prescription for the
growth  and  activity of  central  black  holes  and their  effect  on
suppressing the cooling and star  formation in massive halos.  D07 use
the initial  mass function  (IMF) of Chabrier  (2003), in  contrast to
Croton \etal (2006),  who adopted a Salpeter IMF.  The total number of
galaxies (centrals  plus satellites) in this  catalogue is 25,801,944,
distributed within a total of 14,752,323 dark matter halos.

The  second semi-analytical  model used  in this  paper is  taken from
Bower  \etal  (2006;  hereafter  B06).   This model  uses  the  Durham
semi-analytical model  GALFORM, which is  described in detail  in Cole
\etal  (2000) and  Benson  \etal (2003),  but  has several  additional
features,  including the  formation  and growth  of  black holes,  AGN
feedback,  and disk  instability  (see B06  for  details). This  model
adopts  a Kennicutt  (1983) IMF  with  no correction  for brown  dwarf
stars. The catalogue of model galaxies consists of 24,569,785 galaxies
distributed over 10,957,827 dark matter halos.

The third and final semi-analytical  to be considered in this paper is
that  of Kang  \etal (2005;  hereafter K05).  Unlike the  previous two
models it does  not include AGN feedback. K05  adopted both a Salpeter
IMF and a Scalo IMF, and found that by adjusting model parameters, the
results for the two IMFs are very similar.  The K05 catalogue of model
galaxies has 752,241 entries distributed over 399,983 halos.

All three SAMs share many basic properties. They are all based on dark
matter halo merging trees obtained directly from $N$-body simulations,
Models D07 and B06 are based on the Millennium Run $N$-body simulation
(Springel  \etal 2005b),  a large  dark-matter-only simulation  of the
$\Lambda$CDM   cosmology   with   $2160^3\simeq   1.0078\times10^{10}$
particles in a periodic box of  $500h^{-1}$Mpc on a side.  The mass of
each  particle is  $8.6  \times 10^8\msunh$,  and  the smallest  halos
identified consist of about  20 particles.  The simulation was carried
out  using a  special version  of  the GADGET-2  code (Springel  \etal
2005a).   The  simulation  used  in  K05 was  carried  out  using  the
vectorized  parallel  $\mathrm{P}^3\mathrm{M}$ code  of  Jing \&  Suto
(2002)    with   $512^3$    particles   distributed    over    a   box
$100h^{-1}\mathrm{Mpc}$   on   a   side.    The   particle   mass   is
$6.2\times10^8h^{-1}\Msun$.  To identify  substructures in  the halos,
all three  SAMs used the  routine SUBFIND developed by  Springel \etal
(2001).  Note  that each model  used a slightly different  approach to
construct their halo merger trees,  which may affect the properties of
the resulting galaxies. Details can  be found in the papers presenting
each individual model.  As  is standard in semi-analytical models, all
three  models  considered  here   take  into  account  basic  physical
processes, such  as gas  cooling, star formation,  supernova feedback,
galaxy  mergers, and  chemical enrichment.   Following White  \& Frenk
(1991), they all  define a cooling radius and  compare it with another
critical radius to separate the  static hot halo regime from the rapid
cooling  regime:   in  D07   and  K05  this   is  the   virial  radius
$r_\mathrm{vir}$,  while B06  use  the free-fall  radius $r_{\rm  ff}$
instead. As mentioned above, both  D07 and B06 include AGN feedback to
suppress cooling flows in massive halos, and both are based on the AGN
model   of  Kauffmann   \&  Haehnelt   (2000),  though   the  detailed
implementations are  different (see below). Finally, both  D07 and B06
took into  account the  effects of re-ionization  of the  universe. In
B06, gas cooling is assumed to be completely suppressed in dark matter
halos with virial  velocities below $50 \kms$ at  redshifts below $z =
6$, while in D07, the  effect of photoionization heating is assumed to
reduce    the    gas     fraction    from    the    universal    value
$f_\mathrm{b}^\mathrm{cosmic}$ to
\begin{equation}
  f_\mathrm{b}^\mathrm{halo}(z, M_\mathrm{vir})
= \frac{f_\mathrm{b}^\mathrm{cosmic}}
  {[1+0.26M_\mathrm{F}(z)/M_\mathrm{vir}]^3}\,,
\end{equation}
where $M_{\rm  vir}$ is the virial  mass of the halo  in question, and
$M_\mathrm{F}$ is a filtering mass (e.g., Gnedin 2000).

Note  that   the  K05  model   does  not  include  AGN   feedback  nor
reionization. Nevertheless, as  we show in section~\ref{sec_comparing}
below,  in terms  of the  stellar mass  distributions, the  K05 yields
results  that are  very comparable  to those  of D07  and B06.  In the
following  two  subsections we  highlight  a  few  of the  differences
between the  three SAMs  considered here that  may have  a significant
impact on the outcome of the stellar masses of the model galaxies.

\subsection{Black hole growth and AGN feedback}

D07 follow  the treatment  of AGN activity  described in  Croton \etal
(2006).   Central massive  black  holes grow  through  two modes:  the
\textit{quasar mode} and the \textit{radio mode}.  In the quasar mode,
black holes grow during galaxy  mergers.  The gas mass accreted during
the merger is  assumed to be proportional to the  total cold gas mass,
but with a lower efficiency for smaller mass systems:
\begin{equation}\label{eq:quasar}
\Delta             m_\mathrm{BH,Q}=\frac{f'_\mathrm{BH}m_\mathrm{cold}}
       {1+(280\mathrm{km\ s}^{-1}/V_{\mathrm{vir}})^2}\,
\end{equation}
where  $f'_{BH}=f_{BH}\times  (m_{\mathrm{sat}}/m_{\mathrm{cen}})$ and
$f_\mathrm{BH} \approx  0.03$ is a constant.  Black  hole accretion is
allowed both in major and minor mergers, and the efficiency is assumed
to be proportional to the  mass ratio of the merging galaxies, $m_{\rm
  sat}/m_{\rm cen}$.  In  the radio mode, the AGN  activity is assumed
to be powered by accretion of hot gas onto the central black hole, and
the accretion rate is assumed to be
\begin{equation}\label{eq:radio}
  \dot{m}_\mathrm{BH,R}=\kappa_\mathrm{AGN}\left(\frac{m_\mathrm{BH}}
      {10^8\Msun}\right)\left(\frac{f_\mathrm{hot}}{0.1}\right)
      \left(\frac{V_\mathrm{vir}}{200\mathrm{km\ s}^{-1}}\right)^3\,
\end{equation}
where $m_\mathrm{BH}$ is the  black hole mass, $f_\mathrm{hot}$ is the
fraction of the  total halo mass in  the form of hot gas  and the free
parameter   $\kappa_   \mathrm{AGN}$   is   set  to   be   $6   \times
10^{-6}\Msun\mathrm{yr}^{-1}$.  It  is  assumed  that the  black  hole
growth during  the radio-mode results  in AGN feedback  which strongly
suppresses the cooling of  hot gas in massive halos\footnote{Note that
  the  model does  {\it not}  incorporate any  direct feedback,  be it
  hydrodynamical or  radiative, from the  quasar-mode accretion.}.  As
shown in  Croton \etal  (2006), this radio-mode  AGN feedback  has the
effect that it completely  stops cooling in halos with $V_\mathrm{vir}
\gta 300 \mathrm{km\ s}^{-1}$ between $z=1$ and the present.  In fact,
AGN feedback is considered to provide a physical `explanation' for the
treatment  of  gas  cooling  in  early  semi-analytical  models.   For
instance, Kauffmann et al. (1999) assume that gas cooling is absent in
halos  with  $V_\mathrm{vir} >  350  \mathrm{km\  s}^{-1}$.  K05  also
follow  this  approach, adopting  a  slightly  higher critical  virial
velocity  of  $390  \mathrm{km\,s}^{-1}$.   So although  they  do  not
incorporate AGN feedback as such, they modify the cooling prescription
so that it effectively has a very similar impact.

The treatment  of black hole  growth in B06  is based on  Kauffmann \&
Haehnelt (2000), and the details  can be found in Malbon \etal (2007).
Central  black  holes  are  assumed  to  grow  through  gas  accretion
triggered by both galaxy mergers  and disk instability, and the growth
is controlled  by an efficiency  parameter $F_{\rm BH}$, which  is the
ratio  between the  gas mass  accreted onto  the black  hole  and that
turned into stars during a  starburst.  Note that in B06, AGN feedback
is effective only  in halos where a static  hot atmosphere has formed.
This is  defined to be the case  when the cooling time  is longer than
the free-fall  time.  They  assume that only  in this case  the energy
from the  central black hole can  suppress the cooling  flows and thus
regulate the cooling rate.  Feedback  in this scheme is similar to the
radio  mode considered  in Croton  \etal (2006),  but the  details are
quite different.   B06 simply assume  that the AGN power  prevents gas
from cooling if
\begin{equation}\label{eq:cool}
L_\mathrm{cool} < \epsilon_\mathrm{SMBH}L_\mathrm{Edd}\,,
\end{equation}
independent  of  the  gas   temperature.   In  the  above  expression,
$L_\mathrm{cool}$  is the  cooling luminosity,  and the  available AGN
power is  parameterized as a fraction  $\epsilon_\mathrm{SMBH}$ of the
Eddington luminosity of the central black hole.

\subsection{Starburst model}

All three models considered here include a prescription for starbursts
triggered by major mergers, defined  as the merger between the stellar
bodies of  two galaxies with a  mass ratio larger than  0.3.  During a
major  merger,  all  the  stellar  mass  in  the  two  progenitors  is
transformed into a spheroidal `bulge'  component, while some or all of
the cold  gas is assumed to undergo  a starburst. In K05  and B06, all
the cold  gas is assumed to turn  into bulge stars, while  in D07 they
adopt the implementation of Somerville \etal (2001), only a fraction
\begin{equation}
  e_\mathrm{burst}=\beta_\mathrm{burst}(m_\mathrm{sat}
  /m_\mathrm{cen})^{\alpha_\mathrm{burst}}\,,
\end{equation}
of  the cold  gas  is  converted into  bulge-stars.  Motivated by  the
numerical  simulation   results  of   Cox  \etal  (2004),   D07  chose
$\alpha_\mathrm{burst}=0.7$ and $\beta_\mathrm{burst}=0.56$.

Besides  these  major-merger  induced  starbursts, D07  and  B06  also
include a  prescription for starbursts triggered  by disk instability.
When a  galaxy disk is  sufficiently massive that its  self-gravity is
dominant, it  is assumed to  be unstable to small  perturbations.  The
instability criterion is based on the quantity
\begin{equation}
\epsilon=\frac{V_\mathrm{max}}{(GM_\mathrm{disk}/r_\mathrm{disk})^{1/2}}\,,
\end{equation}
where $V_{\rm  max}$ is the maximum  value of the rotation  curve, and $M_{\rm
  disk}$  and $r_{\rm  disk}$  are the  mass  and scale  length  of the  disk,
respectively.  If, at any  step, $\epsilon < \epsilon_\mathrm{disc}$, the disk
is assumed to  be unstable. In D07, enough stellar mass  is transferred to the
bulge such  that the disk  will restore stability.   While in B06,  the entire
mass  in the  disk will  be transferred  to the  bulge, with  any  gas present
assumed to undergo a starburst (van  den Bosch 1998; Mo \etal 1998; Cole \etal
2000; Croton et al. 2006; Bower et al. 2006).

\section{Mock galaxy redshift surveys and groups}
\label{sec_MGRS}

The  end product  of each  SAM considered  here is  a large  sample of
galaxies  distributed over  the dark  matter  halos in  a large  cubic
simulation box. One approach would  be to compare these galaxy samples
{\it directly} with the SDSS data. However, this ignores the fact that
the  latter is  affected by  observational selection  effects,  and by
inaccuracies related  to our  halo-based group finder.  In particular,
the group finder used to  identify galaxy groups from the SDSS suffers
from incompleteness  and from  contamination by interlopers  (see Yang
\etal 2005a, 2007).  Furthermore, the halo masses for  the SDSS groups
are estimated from the  ranking of the characteristic luminosities and
stellar   masses  of   the   groups,  which   effectively  assumes   a
deterministic  (i.e. zero scatter)  relation between  these quantities
and  halo mass.   In reality  there  will be  non-zero scatter,  which
results in errors in the inferred  halo mass, which are expected to be
larger for  less massive  halos (see Y09b).   To test the  severity of
such effects,  we construct mock galaxy redshift  surveys (MGRSs) from
the D07 and B06 SAM simulation  boxes to which we apply our halo-based
group  finder.   Since   the  box  size  of  the   K05  SAM  is  small
($100\mpch$), we do not construct a MGRS for this model.

Our construction of  the MGRS here is similar to  that described in Li
\etal (2007a;  see also  Yang \etal 2004).   First, we  stack $3\times
3\times  3$ replicates  of  the  simulation box  and  place a  virtual
observer at  the center  of the stacked  boxes.  Next, we  assign each
galaxy ($\alpha$,  $\delta$)-coordinates and remove the  ones that are
outside the mocked  SDSS survey region.  For each  model galaxy in the
survey   region,  we   compute  its   redshift  (which   includes  the
cosmological  redshift due  to the  universal expansion,  the peculiar
velocity, and a $35\kms$ Gaussian line-of-sight velocity dispersion to
mimic  the  redshift  errors  in  the  data),  its  $r$-band  apparent
magnitude (based  on the $r$-band  luminosity of the galaxy),  and its
absolute magnitude  $\rmag$ which is  $K+E$ corrected to  $z=0.1$. For
D07,  where  only  absolute  magnitudes  in  vega  $UBVRI$  bands  are
available,  we convert  them into  SDSS $g$-  and $r$-bands  using the
relations  provided in  Fukugita \etal  (1996). We  eliminate galaxies
that  are   fainter  than  the  SDSS   apparent-magnitude  limit,  and
incorporate   the   position-dependent   incompleteness  by   randomly
eliminating  galaxies according to  the completeness  factors obtained
from the survey  masks provided by the NYU-VAGC  (Blanton \etal 2005).
Finally we construct group catalogues  from the MGRSs for both D07 and
B06, using the same halo-based group  finder as used for the real SDSS
DR4.

Note that in the MGRSs, stellar masses are obtained directly from the SAMs,
i.e., they are not obtained from their $g$- and $r$-band magnitudes using
Eq.~\ref{eq:MtoL}, as for the SDSS galaxies. The reason is that in SAMs,
photometric properties of galaxies are calculated using stellar population
synthesis models, using stellar masses taken from these derived magnitudes
will introduce addtional uncertainties.  For the halo masses, however, we do
not use the actual masses of the halos in the SAMs. Rather, we assign each
mock group a halo mass based on the ranking of its characteristic stellar
masses, in the same way as we assigned halo masses to the SDSS groups.

\section{The stellar mass components of galaxy groups}
\label{sec_comparing}

In what follows we compare the predictions of the SAMs described above
with the observational results.  The quantities to be compared include
the global stellar mass functions,  the total stellar mass in halos of
different masses, and the conditional stellar mass functions.

\subsection{The galaxy stellar mass functions:
central versus satellite galaxies}

\begin{figure*} \epsscale{0.9}
\plotone{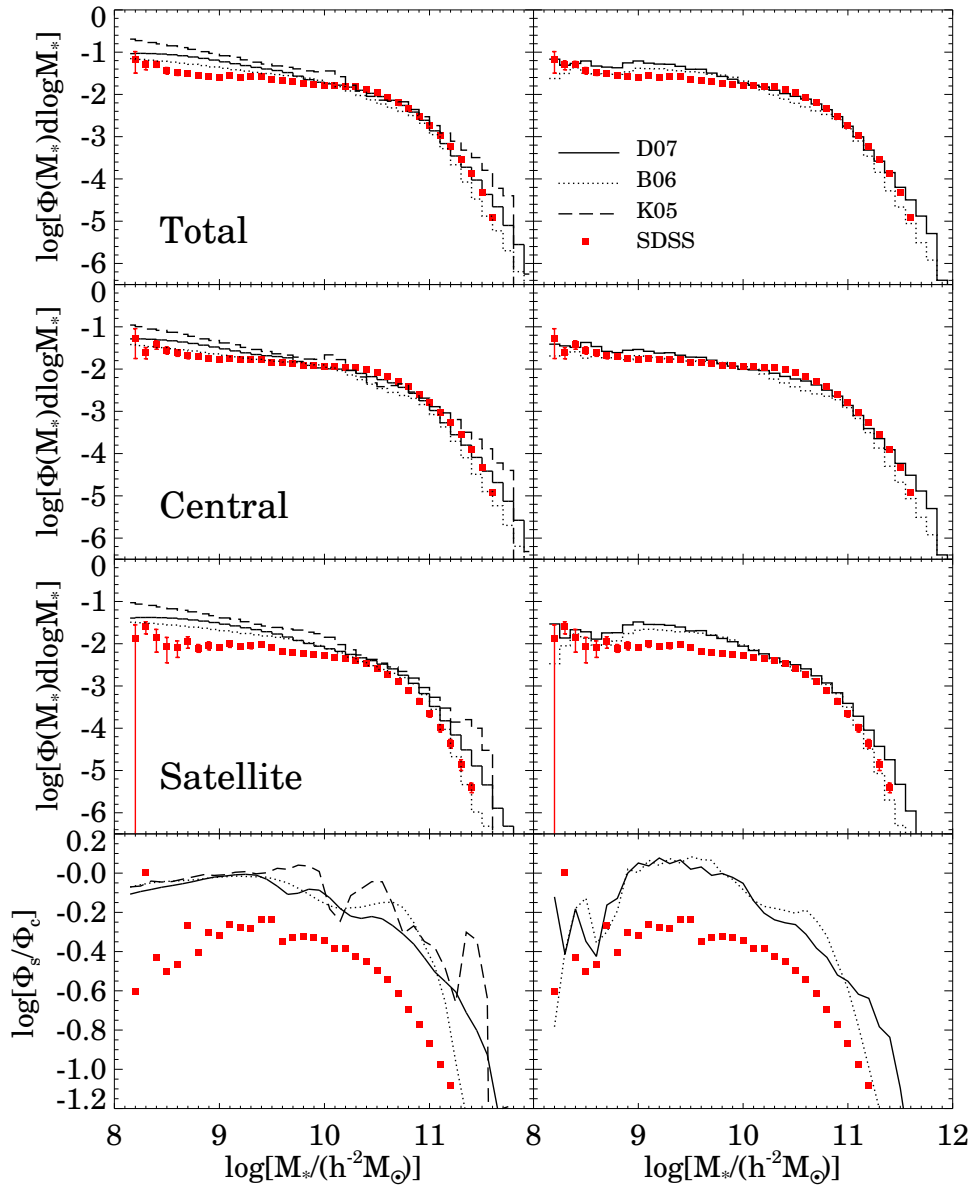}
\caption{Galaxy  stellar mass  functions.  Results shown  in the  left
  panels are obtained directly from the SAMs in the cubic boxes, while
  the right panels are obtained  from the galaxy groups extracted from
  the MGRSs. The  first three rows on top  show stellar mass functions
  for all,  central and satellite galaxies,  respectively.  The solid,
  dotted  and   dashed  histograms  are  results   obtained  from  the
  semi-analytical models of D07,  B06 and K05, respectively. Since the
  statistical errors for  SAMs are negligible, we do  not plot them in
  the figure.   For comparison, symbols  with error bars are  the SDSS
  observational data obtained  by Y09b.  In the bottom  row of panels,
  we show the  ratio of $\Phi_s/\Phi_c$ (third v.s.  second panel) for
  the  corresponding   SAMs  (lines)  and   SDSS  observation  (dots),
  respectively.}
\label{fig:mf}
\end{figure*}

We start by comparing the  stellar mass functions for all, central and
satellite galaxies. The results  are shown in Fig.~\ref{fig:mf}, where
the symbols with errorbars indicate  the SDSS data, and the histograms
correspond to the different SAMs, as indicated in the upper right-hand
panel. The  left-hand panels show the model  results obtained directly
from the cubic SAM simulation  boxes, while the right-hand panels show
the results obtained from groups selected from the MGRSs (only for D07
and B06).

Comparing  the   results  obtained  directly  from   the  cubic  boxes
(left-hand panels)  with those from the MGRSs  (right-hand panels), it
is evident that the survey selection effects and the contamination due
to   the   group  finder   do   not   change   any  of   the   results
qualitatively.  Some small differences  are apparent  at the  low mass
end,  which partially  reflects  the fact  that  the effective  survey
volume is small and cosmic variance is large.

From the  upper left-hand panel  of Fig.\ref{fig:mf} it is  clear that
the stellar mass functions of galaxies predicted by the three SAMs are
roughly  consistent with  the data  in the  intermediate  stellar mass
range  ($\log  \Mstar  \sim  10.2-11.0$). However,  all  three  models
significantly over-predict  the stellar mass function  at the low-mass
end.   At the  high-mass  end ($\log\Mstar  \ga  11.0$), D07  slightly
over-predicts,  B06  slightly  under-predicts, and  K05  significantly
over-predicts the stellar mass function.

In order  to examine the  discrepancies between model  predictions and
data  in more  detail, we  next  consider the  stellar mass  functions
separately for central and satellite  galaxies. As shown in the panels
in  the second  row of  Fig.~\ref{fig:mf}, the  B06 model  is  in good
agreement with the  observed stellar mass function of  centrals at the
low-mass end, but significantly  under-estimates the number density of
massive  centrals (those  with $\log  \Mstar\ga 10.0$).   D07 slightly
over-predicts the number of central  galaxies at the low-mass end, but
fairs well  at the high-mass  end.  K05, finally, only  reproduces the
data in the intermediate mass  range. For satellite galaxies (shown in
the  panels in  the third  row of  Fig.~\ref{fig:mf}), all  three SAMs
over-predict the  number density of satellite  galaxies, especially at
the low-mass  end. The B06  model fairs best, and  actually reproduces
the number  densities of massive  satellites, but overall it  is clear
that the SAMs predict too many satellite galaxies. This is illustrated
even more  clearly in  the lower row  of panels  of Fig.~\ref{fig:mf},
which show the  ratio $\Phi_{\rm s}/\Phi_{\rm c}$ of  the stellar mass
functions  of   satellite  and  central  galaxies.    All  three  SAMs
over-predict this ratio by about  a factor of two. The implications of
these results are discussed in Section \ref{sec_missing}.

\subsection{The stellar masses of central and satellite galaxies
in halos of different masses}\label{sec:c_s}

\begin{deluxetable}{lcccc}
  \tabletypesize{\scriptsize}  \tablecaption{The  best fit  parameters
    for  the   stellar  mass  function  of   central  group  galaxies}
  \tablewidth{0pt}  \startdata  \cline{1-5}\\ Source  &  $\log M_0$  &
  $\log  M_1$ &  $\alpha$ &  $\beta$\\ \cline{1-5}\\  Y09b &  10.306 &
  11.040 & 0.315 & 4.543 \\ De Lucia \& Blaizot (2007; D07) & 10.128 &
  11.300 & 0.451 & 2.164\\ Bower et al. (2006; B06) & 9.967 & 11.493 &
  0.466 & 1.844 \\ Kang et al. (2005; K05) & 10.494 & 11.719 & 0.401 &
  1.392 \\  \enddata \tablecomments{$M_0$ is in  units of $h^{-2}\Msun$
    and $M_1$ in $h^{-1}\Msun$}
\label{tab:fit}
\end{deluxetable}

\begin{figure*}  \epsscale{1.0}
\plotone{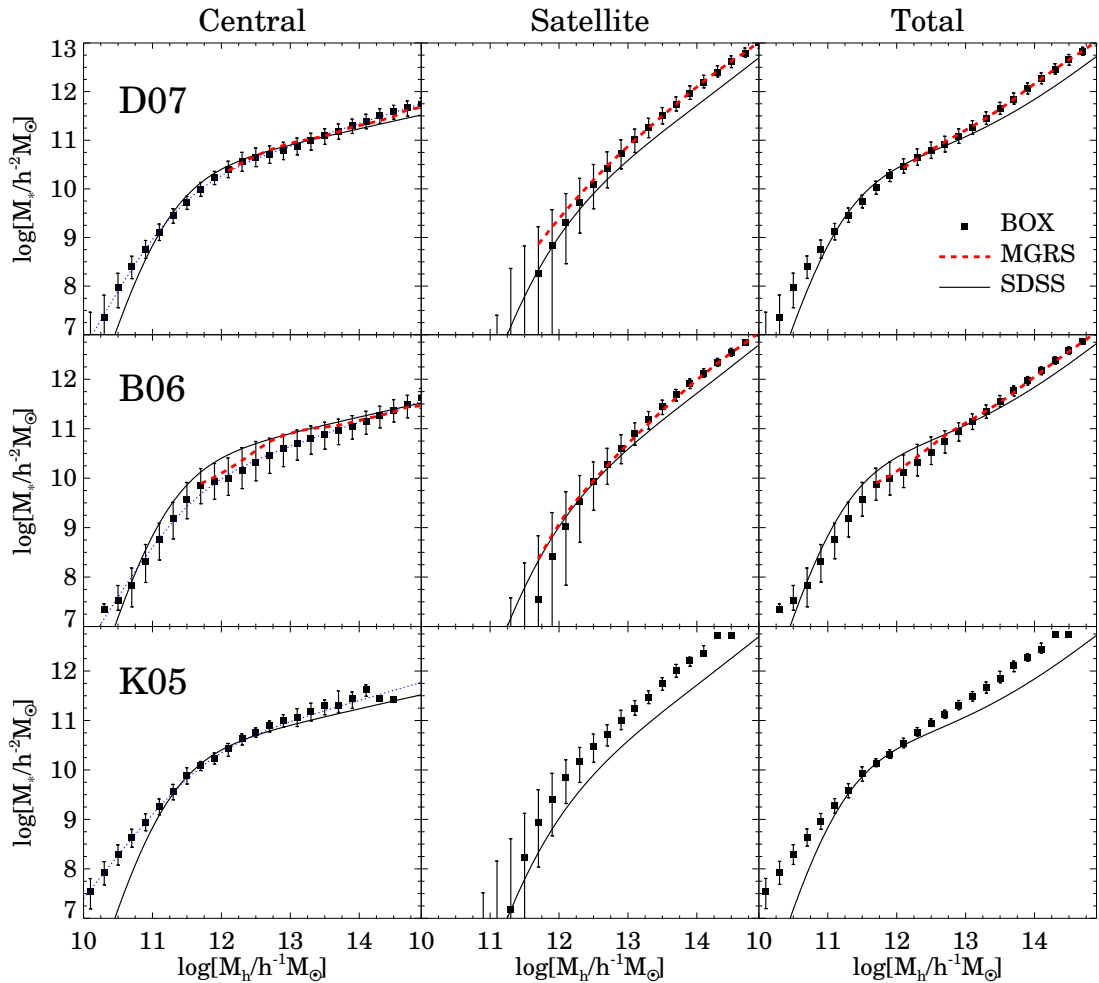}
\caption{The  total  stellar  masses  of central,  satellite  and  all
  galaxies in  halos of different  masses.  Results from D07,  B06 and
  K05 are shown  as symbols with error bars, in  the upper, middle and
  lower  row  panels,  respectively.   The  error bars  in  this  plot
  indicate  the 84\% confidence  levels in  different halo  mass bins.
  For   comparison,  in  each   panel,  the   solid  line   shows  the
  observational results  obtained by Y09b from SDSS.  The dotted lines
  shown in  the left column  panels are the  best fit results  for the
  SAMs.  In the upper two  rows of panels, results obtained from MGRSs
  (i.e.~based   on    extracted   groups)   are    shown   as   dashed
  lines.}\label{fig:halo_s}
\end{figure*}

In  order to look  into the  origin of  the discrepancy  between model
prediction  and  observation,  Fig.\ref{fig:halo_s}  shows  the  total
stellar masses  of central  galaxies (left panel),  satellite galaxies
(middle panels), and all galaxies  (right panels), as functions of the
host halo mass.  The results obtained directly from the SAM simulation
boxes (BOX) are shown as solid squares with error bars (which indicate
the corresponding  84\% confidence  levels).  For comparison,  we also
show,  in the  upper two  rows, the  results obtained  from  the MGRSs
(dashed  lines). In general,  the results  obtained directly  from the
simulation boxes agree  well with those obtained from  the MGRS.  This
is extremely reassuring, as it  implies that the results obtained from
the Y07 galaxy group catalogue are reliable.

Let us  first consider  the central galaxies.   As shown in  Y09b, the
conditional  probability  distribution,  $\calPc(\Mstar|\Mh)$, that  a
halo  of mass $\Mh$  hosts a  {\it central}  galaxy with  stellar mass
$\Mstar$ can be  described by a log-normal distribution  with the mean
given by
\begin{equation}
\label{eq:halo_s}
\langle M_{\ast,{\rm c}}\rangle(\Mh) = M_0
\frac{(\Mh/M_1)^{\alpha+\beta}}{(1+\Mh/M_1)^\beta}\,.
\end{equation}
We use this  relation to fit the stellar mass -  halo mass relation of
centrals predicted  by the  three SAMs.  The  best fit  parameters are
listed in Table  \ref{tab:fit}, and the results are  also shown in the
left  columns of  Fig.  \ref{fig:halo_s} as  the  dotted lines.  These
results should be compared to the best fit values obtained by Y09b for
central galaxies in  the SDSS group catalogue, shown  in each panel as
the solid line: the corresponding  best fit parameters are also listed
in Table \ref{tab:fit}.  Note that the predictions of  D07 and K05 for
the stellar mass  - halo mass relation are  quite similar. Both models
slightly over-predict the stellar  mass of central galaxies in massive
halos  with  $\Mh\ga  10^{14}\msunh$,  and severely  over-predict  the
stellar  masses in low  mass halos  ($\Mh\la 10^{11}\msunh$).   In the
intermediate halo-mass range, both D07 and K05 match the observational
data reasonably well. The B06 model  yields a stellar mass - halo mass
relation  of central  galaxies that  is quite  different:  although it
matches the data at both  the massive ($\Mh \sim 10^{14.5}\msunh$) and
low-mass   ($\Mh   \sim   10^{11}\msunh$)   ends,   it   significantly
under-predicts the stellar masses in intermediate-mass halos.

It is instructive  to look at the best fit  parameters listed in Table
\ref{tab:fit}.  As one can  see, B06  predicts a  lower characteristic
stellar  mass, $M_0$,  than the  other two  models, which  reflects an
overall lower  amplitude of the stellar  mass - halo  mass relation of
central galaxies.  Compared to  the SDSS data, the biggest discrepancy
for all  the three SAMs concerns  the slope, $\beta$,  at the low-mass
end.  The  models predict $1.4 \la  \beta \la 2.2$,  much smaller than
$\beta\sim 4.5$ obtained from the  SDSS.  This indicates that the star
formation efficiency  in low-mass halos increases with  halo mass much
faster than assumed in the  SAMs.  This is consistent with the results
obtained  by  Mo \etal  (2005),  who found  that  it  is difficult  to
reproduce the  observed low-mass end  of the stellar mass  function if
the star formation in low-mass  halos is mainly regulated by supernova
feedback, and  by Pasquali \etal  (2009b), who have  demonstrated that
typical SAMs significantly over-predict the stellar population ages of
central galaxies in low mass halos.

Next we focus on the contribution of satellite galaxies to the stellar
mass budget.  The panels in the middle column of Fig.~\ref{fig:halo_s}
show  the  total  stellar  mass  contained in  satellite  galaxies  as
function of halo mass, defined as the sum of the stellar masses of all
the member satellite galaxies with $\Mstar \geq 10^8h^{-2}\Msun$.  The
solid  line in  each of  the middle-column  panels shows  the observed
average,  obtained  by Y09a,  properly  converted  to this  particular
stellar  mass  limit of  $10^8h^{-2}\Msun$.  Clearly,  all three  SAMs
significantly  over-predict  the   total  stellar  mass  contained  in
satellites, especially in halos with $\Mh \ga 10^{13}\msunh$.

Finally, the right-hand panels of Fig.~\ref{fig:halo_s} show the total
stellar  mass   (central  plus   all  satellites  with   $\Mstar  \geq
10^8h^{-2}\Msun$) as function of halo mass. Compared to the SDSS data,
D07  and K05 significantly  over-predict the  total stellar  masses in
halos at  both the high-  and low-mass ends.   On the other  hand, B06
also  over-predicts  the total  stellar  mass  in  massive halos,  but
under-predicts the total stellar mass in halos with $10^{11}\msunh \la
\Mh \la  10^{12.5}\msunh$.  It  is clear that  the discrepancy  at the
low-mass end  is due to centrals,  while that at the  high-mass end is
due to satellites.

\subsection{The conditional stellar mass functions}
\label{sec:CSMF}

\begin{figure*} \epsscale{1.0}
\plotone{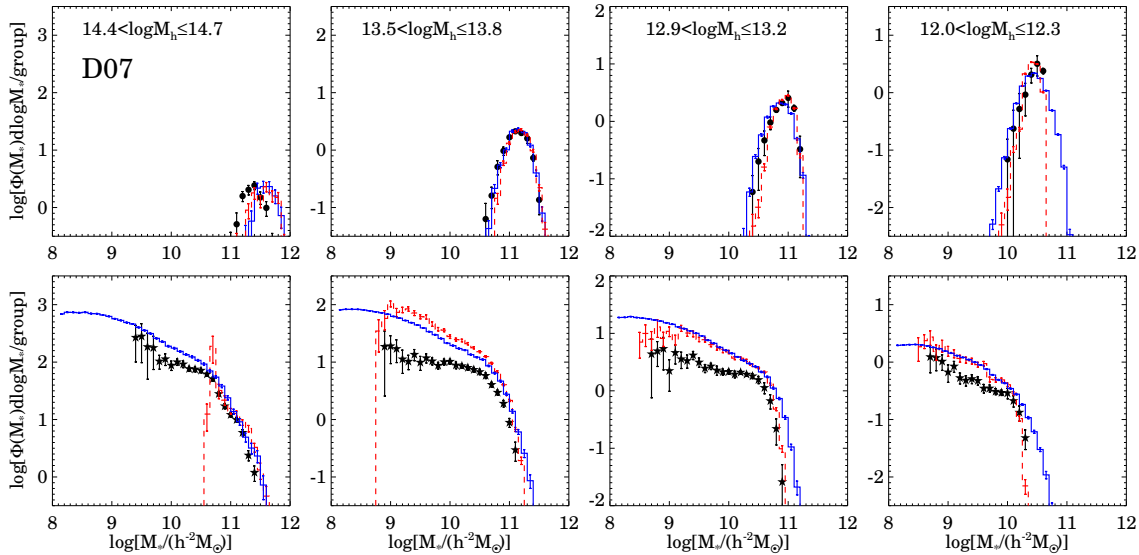}
\caption{The conditional stellar mass functions (CSMFs) of galaxies in
  halos of different mass bins.  Shown in the upper panels are results
  for  central  galaxies, while  the  lower  panels  give results  for
  satellite   galaxies.   Symbols   with  error   are  for   the  SDSS
  observational results,  whereas the histograms with  error bars show
  the SAM predictions of D07.   The blue solid histograms are obtained
  directly for the  raw models in a cubic box,  whereas the red dashed
  ones are obtained for groups extracted for MGRSs. The error bars for
  the SAM predictions are obtained using 500 bootstrap resamplings.}
\label{fig:csmf_DL}
\end{figure*}

\begin{figure*} \epsscale{1.0}
\plotone{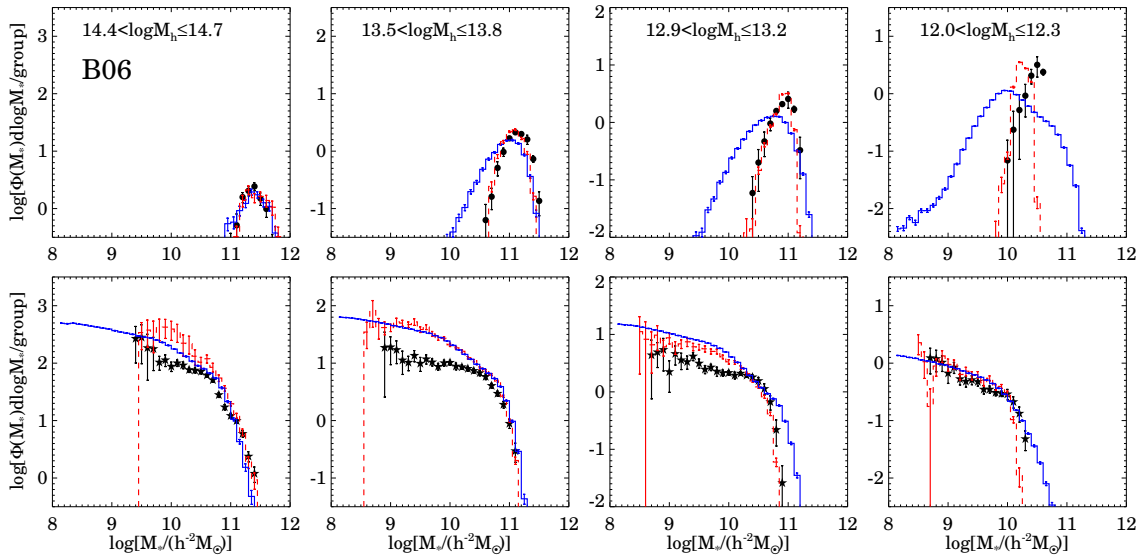}
\caption{Similar to Fig.~\ref{fig:csmf_DL} but for the SAM predictions
  of B06.}
\label{fig:csmf_bower}
\end{figure*}

\begin{figure*} \epsscale{1.0}
\plotone{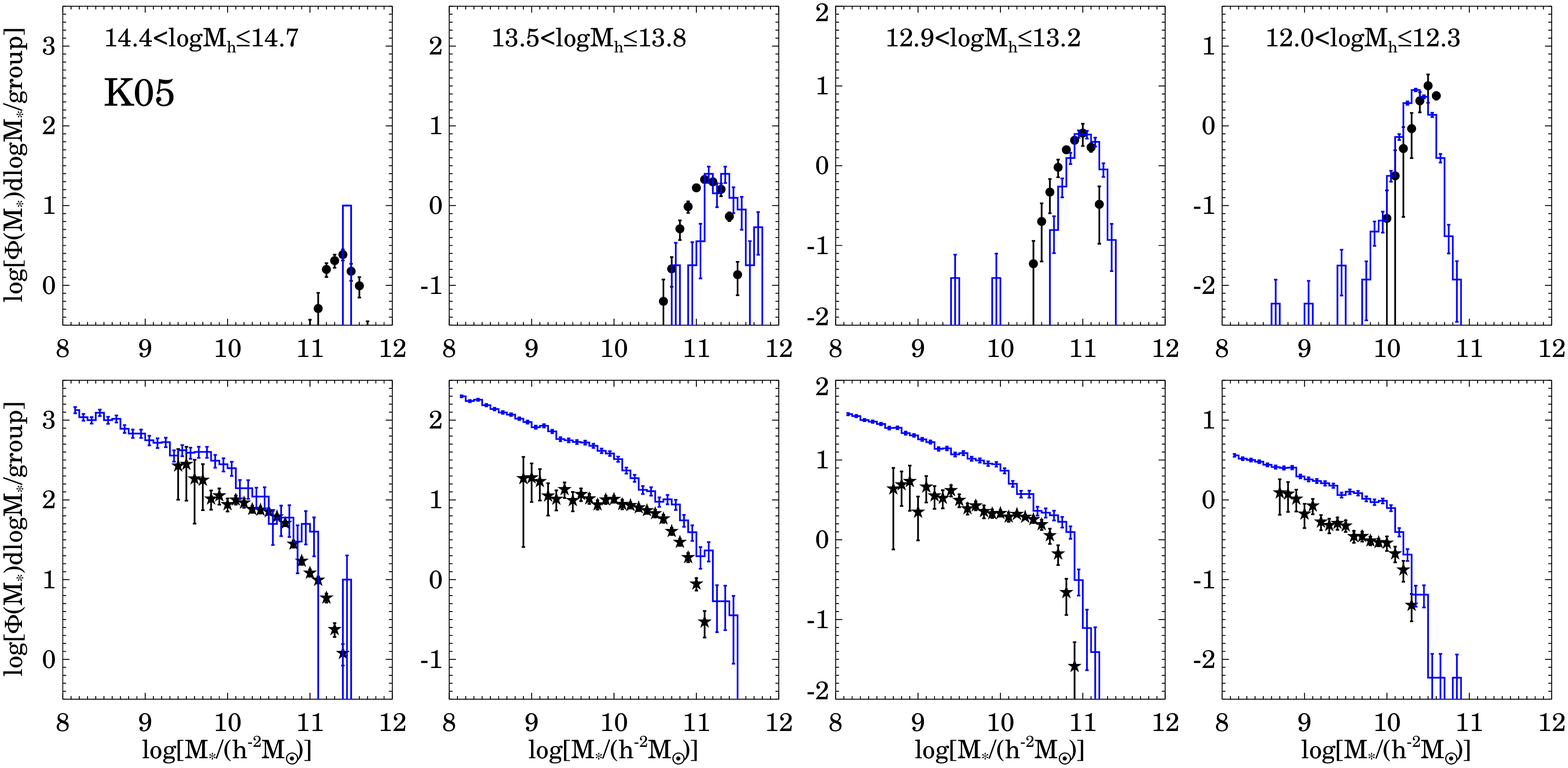}
\caption{Similar to Fig.~\ref{fig:csmf_DL} but for the SAM predictions
  of K05. Note  that we do not  have results based on a  MGRS for this
  model. }
\label{fig:csmf_kang}
\end{figure*}

To get more insight regarding  the origin of the discrepancies between
model  predictions and data,  we now  examine the  conditional stellar
mass function (CSMF),  $\Phi(\Mstar|\Mh)$, which describes the average
number of galaxies  of stellar mass $\Mstar$ that reside  in a halo of
mass $\Mh$. The  CSMF has been measured directly  from the SDSS galaxy
group  catalogue by  Y09b.  For  the  three SAMs  considered here,  we
determine  the CSMFs  separately  for central  and satellite  galaxies
directly from the simulation boxes, and  for B06 and D07 also from the
group  catalogues  constructed from  their  corresponding MGRSs.   The
results of  the direct  measurements are shown  as the solid  lines in
Figs.~\ref{fig:csmf_DL},  \ref{fig:csmf_bower} and \ref{fig:csmf_kang}
for D07, B06 and K05, respectively, while those obtained from the MGRS
are    shown    as    dashed   lines    (in    Figs.~\ref{fig:csmf_DL}
and~\ref{fig:csmf_bower} only).  In each panel, the symbols with error
bars indicate the observational results of Y09b.

Let us first  focus on the CSMFs for central galaxies.  In D07, we see
that  the  mock  group  results   agree  very  well  with  the  direct
measurements,  although the  former  yield a  width  that is  somewhat
smaller  than the true  width, obtained  directly from  the simulation
boxes, in low mass halos. In the case of the B06 model, however, there
is a  huge discrepancy between  the CSMF of central  galaxies obtained
from  the mock  group  results  and that  obtained  directly from  the
simulation  boxes. In  particular, the  width  of the  former is  much
smaller than that of the latter, indicating that the width of the CSMF
obtained  from our SDSS  galaxy group  catalogue may  be substantially
underestimated. As discussed in Y09b, this is due to the fact that the
assignment of  halo masses to the  groups assumes zero  scatter in the
$\Mstar$-$\Mh$ relation. If  the true $\Mstar$-$\Mh$ relation contains
a large  amount of  scatter, as is  the case  for the B06  model, this
results  in significant  errors in  the  assigned halo  masses, and  a
significant  underestimate of  the  width of  the  CSMF for  centrals.
Hence, the width  obtained from the galaxy group  catalogues has to be
considered a lower limit on the true width.

It   is   clear   from   a   comparison   of   Figs~\ref{fig:csmf_DL},
\ref{fig:csmf_bower}  and  \ref{fig:csmf_kang}   that  the  B06  model
predicts  a much  larger  scatter  in the  stellar  masses of  central
galaxies  than  the  D07  and  K05  models,  especially  in  low  mass
halos\footnote{Although the origin of  this dramatic difference is not
  entirely clear to us, we believe that it originates from the way B06
  implement  AGN  feedback.}.  It  is therefore  important  to  obtain
constraints  on  the true  amount  of  scatter  in the  $\Mstar$-$\Mh$
relation of central galaxies.  It should be clear from the above, that
we cannot use the SDSS group  catalogue for this. After all, we obtain
similar  widths for  the CSMFs  of centrals  when we  apply  our group
finder to  the D07 and B06  MGRSs, even though  their intrinsic widths
are clearly  very different. However, recently More  \etal (2009) have
constrained the scatter, $\sigma_{\log\Mstar}$,  as a function of halo
mass using  satellite kinematics. Their method, which  has been tested
using   detailed   mock  galaxy   redshift   surveys,  indicate   that
$\sigma_{\log\Mstar} \sim  0.16$~dex. A similar amount  of scatter was
inferred   by  Cacciato  \etal   (2009)  from   an  analysis   of  the
galaxy-galaxy lensing signal in the  SDSS, as measured by Seljak \etal
(2005)  and  Mandelbaum  \etal  (2006).   Taking  these  observational
constraints at  face value, they clearly  rule out the  huge amount of
scatter predicted by  the B06 model.  Interestingly, though,  it is in
excellent agreement with the amounts  of scatter predicted by both the
D07 and K05 models.

Focusing  on the  peak positions  of  the CSMFs  of central  galaxies,
rather  than the  scatter,  we  notice the  same  discrepancies as  in
Fig.~\ref{fig:halo_s}: the  D07 and  K05 models over-predict  the mean
stellar  mass  of  central   galaxies  in  massive  halos,  while  B06
under-predicts the mean stellar mass of central galaxies in halos with
$\Mh \sim 10^{12} h^{-1} \Msun$.

For  satellite   galaxies,  the  overall  mock   group  results  agree
reasonably  well  with  the  direct  measurements\footnote{The  abrupt
  truncation in  the CSMF obtained from  the D07 MGRS  apparent in the
  lower-left  panel of  Fig.~\ref{fig:csmf_DL} is  a  manifestation of
  cosmic  variance.}.  However,  the  CSMF  for  the  mock  groups  in
relatively massive halos is  slightly overestimated, and in relatively
small  halos   slightly  underestimated  especially   at  the  massive
end. Note  that, in a  mock group, the  most massive galaxy  is always
defined to  be the central galaxy.   Thus, if the true  central is not
the most  massive one (i.e.  one  or more satellites  are more massive
than  the  central),  the  mass  of  the  most  massive  satellite  is
underestimated, leading to an underestimate of the conditional stellar
mass function  at the  massive end.  Furthermore,  the masses  of some
mock  groups may  be over-estimated  by the  ranking of  their stellar
masses if  their stellar masses  are exceptionally large.   Because of
these  uncertainties, it  is perhaps  more meaningful  to  compare the
observational results with the results  obtained from the MGRS. As one
can see, all the SAMs over-predict the CSMF of satellites at the faint
end, especially in relatively massive halos.

\section{What is missing in semi-analytical models?}
\label{sec_missing}

The model-data comparisons presented in the previous section show that
all three  SAMs considered fail  in the following two  aspects. First,
the models  over-predict the number  of satellite galaxies by  about a
factor of two, for halos  of all masses. Second, the predicted stellar
mass -  halo mass relation of  central galaxies is too  shallow at the
low-mass end. In this section we investigate how these problems may be
remedied.

In all three SAMs considered above, once  a halo is accreted by a larger halo,
its galaxies (now  `satellite galaxies') are assumed to  either merge with the
central galaxy of the host halo  or to remain as individual satellite galaxies
in their new host  halo (but see a recent attempt by  Henriques \& Thomas 2009
who  have introduced  into  the SAM  by  D07 the  tidal  stripping of  stellar
material from satellite galaxies during mergers). Other possibilities, such as
satellite-satellite merging,  and the  stripping and disruption  of satellites
due to tidal forces, are not  taken into account.  However, both processes are
believed to play an important role.  In fact, numerous studies in recent years
have argued that reconciling halo occupation statistics with halo merger rates
requires that a  significant fraction of satellite galaxies  is indeed tidally
disrupted (e.g., Conroy, Ho \&  White 2007; Conroy, Wechsler \& Kravtsov 2007;
Kang \& van  den Bosch 2008; Y09a). In addition, Kim  \etal (2009) have argued
that  both tidal disruption  and satellite-satellite  merging are  required in
order to  reproduce the  two-point correlation function  of galaxies  on small
scales.  Since  satellite-satellite merging does not reduce  the total stellar
mass in the  host halo, and so cannot alleviate  the discrepancy between model
and data shown  in the right panels of  Fig.~\ref{fig:halo_s}, in what follows
we focus on the impact of tidal stripping.
\begin{figure} \epsscale{1.0}
\plotone{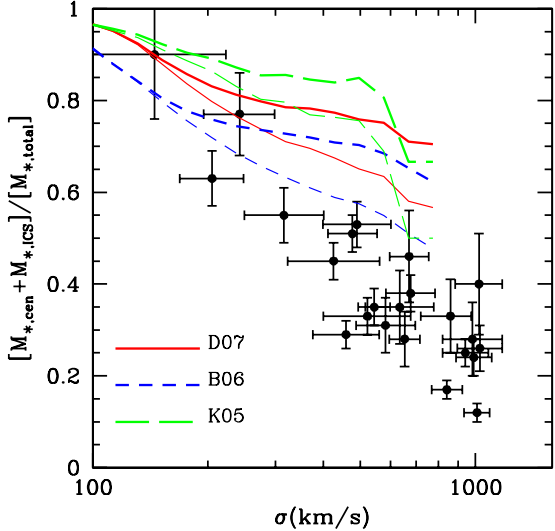}
\caption{The  fraction of  the  total stellar  mass  contained in  the
  central galaxy and  in intra-cluster stars as a  function of cluster
  velocity dispersion. Data with error bars are obtained from Gonzalez
  et  al.~(2007), measured  within  $r_{500}$ for  the  23 groups  and
  clusters in  their sample.  The solid, dashed  and long-dashed lines
  are model predictions for D07,  B06 and K05, respectively. The thick
  and thin lines correspond to  the two extreme cases we analyzed (see
  text for details).}
\label{fig:ICS}
\end{figure}

As shown in the panels  in the middle column of Fig.~\ref{fig:halo_s},
all three SAMs over-predict the {\it total} stellar mass of satellites
in halos with $M_h\ga 10^{13.0}\msunh$.  In order to reduce the amount
of stellar mass  locked up in satellite galaxies  without changing the
star formation  prescription in the models, a  significant fraction of
the  stars in  satellite  galaxies has  to  be stripped  and form  the
stellar halos  and intra-cluster stars  (ICS)\footnote{Throughout this
  paper  we  will use  the  terms  `ICS'  and `stellar  halo'  without
  distinction.}   observed  in  our  Milky  Way halo,  in  groups  and
clusters,  and  around  isolated  galaxies (e.g.,  Helmi  \etal  1999;
Ferguson  \etal 2002; Yanny  \etal 2003;  Zibetti, White  \& Brinkmann
2004; Gonzalez,  Zabludoff \& Zaritsky  2005, 2007; Seigar,  Graham \&
Jerjen 2007; Bell  \etal 2008). In addition, recent hydrodynamical
simulations also suggest a very significant fraction of ICS in clusters
of galaxies. (Puchwein et al. 2010).
As a quantitative  measure of how many
stars are observed  in the form of ICS, the  solid dots with errorbars
in Fig.~\ref{fig:ICS} show the  observed fraction of the total stellar
mass present in  groups and clusters that is  contained in the central
galaxy  and  the ICS  as  a  function  of the  line-of-sight  velocity
dispersion  of member galaxies.   These data  are taken  from Gonzalez
\etal (2007),  under the  assumption that satellite  galaxies, central
galaxies and  the ICS all  have a similar stellar  mass-to-light ratio
(in  the  $i$-band).   We  compare  the observational  data  with  the
predictions of the  three SAMs: D07 (solid lines),  B06 (dashed lines)
and  K05 (long-dashed  lines ),  assuming  that the  entire excess  of
stellar mass in satellite galaxies  predicted by the models resides in
ICS.  Here  the halo mass  is converted into a  line-of-sight velocity
dispersion using  equation (6)  in Yang \etal  (2007).  Note  that the
data  of Gonzalez  \etal  (2007) are  obtained  within $r_{500}$,  the
radius  within which  the cluster  mass density  exceeds  the critical
value by a factor of 500.   This factor has been taken into account in
all the  three SAM predictions:  the total mass of  satellite galaxies
within  $r_{500}$ is estimated  by assuming  that the  distribution of
satellite  galaxies follows  the NFW  (Navarro, Frenk  \&  White 1997)
profile  with   concentration  appropriate   for  the  halo   mass  in
question. For the distribution of  ICS, we consider two cases.  Case I
assumes  that the  ICS have  the  same distribution  as the  satellite
galaxies,    and   the    corresponding   results    are    shown   in
Fig.~\ref{fig:ICS} as the three thin  lines.  Case II assumes that all
ICS are  distributed within  $r_{500}$, and the  corresponding results
are  shown in  Fig.~\ref{fig:ICS}  by the  three  thick lines.  Except
perhaps for  the Case I scenario  of B06, all  model predictions thus
obtained significantly  over-predict the ratio  $[M_{\ast,{\rm cen}} +
  M_{\ast,{\rm ICS}}]  / M_{\ast,{\rm total}}$ in  massive halos. This
suggests that all models predict too many stars, and that a consistent
result cannot  be obtained without changing the  prescription for star
formation.  This  is  further   corroborated  by  the  fact  that  the
observational estimates by Gonzalez~\etal (2007) for the ICS component
lie  at the  upper  end of  the  reported values,  with other  studies
finding considerably lower numbers  (e.g. Zibetti, White \& Brinkmann,
2004).

As  one can see  from Figs.~\ref{fig:csmf_DL}  to \ref{fig:csmf_kang},
the main problem  arises because the models predict  too many low-mass
satellite  galaxies,  especially in  relatively  massive halos.  Since
satellite galaxies were central  galaxies before they were accreted by
their host  halos, this discrepancy  suggests that the  star formation
efficiency in low-mass halos needs to  be reduced in the SAMs. This is
further supported  by the  fact that the  SAMs predict a  low-mass end
slope $\beta$  for the  stellar mass -  halo mass relation  of central
galaxies that  is much too shallow (see  section~\ref{sec:c_s}) and by
the fact  that SAMs  typically over-predict the  mass-weighted stellar
ages  of low-mass  galaxies (Pasquali  \etal 2009b).   Furthermore, as
recently demonstrated in Y09a, if one assumes that the central stellar
mass -  halo mass relation is  independent of redshift  (see also Wang
\etal 2006), then a model that can reproduce the observed stellar mass
- halo  mass relation  of  centrals can  also  reproduce the  observed
stellar mass function of satellite galaxies and the observed ICS.  All
these  results  clearly suggest  that  current semi-analytical  models
over-predict the star formation efficiencies in low mass halos.

\section{Summary}\label{sec_sum}

We have evaluated three SAMs by De Lucia \& Blaizot (2007; D07), Bower
\etal  (2006; B06)  and Kang  \etal (2005;  K05),  using observational
measurements of the stellar mass  functions, the total stellar mass in
halos of different masses,  and the conditional stellar mass functions
for central and satellite galaxies.   Our results can be summarized as
follows.
\begin{itemize}
\item All  three SAMs  predict stellar mass  - halo mass  relations of
  central  galaxies  that  have  a  similar  shape,  but  a  different
  normalization. While D07 and  K05 over-predict the stellar masses of
  centrals  in  massive and  low  mass  halos,  B06 under-predict  the
  stellar masses of  centrals in intermediate mass halos.  None of the
  SAMs reproduces  the observed  steep slope of  this relation  at the
  low-mass end.
\item  All three  SAMs  over-predict  the ratio  of  the stellar  mass
  functions     of     satellites     and     centrals,     $\Phi_{\rm
    s}(\Mstar)/\Phi_{\rm c}(\Mstar)$, by about  a factor of two at all
  $\Mstar$.
\item All three SAMs over-predict the total stellar mass and number of
  low-mass satellite galaxies, especially in massive halos.
\end{itemize}

Neither  of the  three SAMs  considered  here takes  account of  tidal
stripping of (the stellar  components of) satellite galaxies.  Rather,
satellite galaxies either survive to  the present day with roughly the
same stellar mass as they has at their epoch of accretion, or they are
accreted by the central galaxy in their host halo. Whether a satellite
survives or is  accreted is determined by the  efficiency of dynamical
friction, which  controls the rate  at which the satellite  loses its
momentum.  In reality,  however,  satellite galaxies  are expected  to
experience tidal  stripping, which not only  increases their dynamical
friction times,  but also may  cause satellite galaxies to  be tidally
disrupted. In either case, a  certain fraction of the stars originally
associated  with satellite galaxies  ends up  being associated  with a
stellar halo (called `intra-cluster stars'  in the case where the host
halo is  cluster-sized). Hence, one  can improve the problem  with the
overproduction of stellar mass  associated with satellite galaxies, by
assuming  that a certain  fraction of  satellite galaxies  are tidally
stripped  or disrupted.  However, we  have  shown that  the amount  of
stripping  required  in  the  SAMs  in order  to  match  the  observed
conditional  stellar mass function  of satellites  would significantly
over-predict the  mass of the ICS-component in  massive halos compared
with  observation. This indicates  that the  problem with  the stellar
masses of  satellite galaxies is not  solely a consequence  of how the
models  treat the evolution  of satellite  galaxies. Rather,  we argue
that the satellite  galaxies in the SAMs are  too massive because they
were already  too massive at their  time of accretion,  when they were
still centrals.  This is  supported by the  fact that the  models also
fail to  reproduce the low-mass end  slope of the stellar  mass - halo
mass   relation  of   centrals.  We   therefore  conclude   that  SAMs
over-predict the star formation efficiencies in low mass halos.

The  main mechanism  that is  invoked in  (semi-analytical)  models of
galaxy formation in order to regulate the star formation efficiency in
low mass halos is supernova  (SN) feedback. Hence, a naive solution to
the  problems identified  here  seems  to be  to  simply increase  the
supernova  feedback efficiency,  typically expressed  in terms  of the
fraction of supernova energy used  to either expel or reheat cold gas.
However, often this efficiency  is already taken to be unrealistically
high. For example,  in the B06 model it is larger  than unity in halos
with a circular  velocity less than $200 \kms$  (see Benson \etal 2003
for a detailed discussion). Another problem with supernova feedback as
the only  mechanism to  suppress star formation  in low mass  halos is
that it requires  star formation, which in turn  requires high surface
densities  of gas.  As shown  in Mo  \etal (2005),  the cold  gas mass
fractions  required  in order  to  maintain  the  needed level  of  SN
feedback are too  high compared to observation.In addition, strong
supernova feedback can lead to overly low effective metal yields,
resulting in metallicities in low mass galaxies that are too low.
Hence,  it is unlikely
that a  mere modification  of SN feedback  can solve the  stellar mass
problems  identified in  this paper,  suggesting that  other  forms of
feedback, perhaps from AGN in  the quasar mode (Di Matteo, Springel \&
Hernquist  2005), are  needed.  An  alternative solution  is  that the
intergalactic medium  (IGM) is preheated  so that the total  amount of
gas that  can be  accreted by a  low mass  halo for star  formation is
reduced (Mo  \& Mao 2002). However,  at the moment there  is no direct
evidence to  support the  idea that the  (high-redshift) IGM  may have
been  substantially pre-heated.  We therefore  conclude that  we still
lack  a proper  understanding  of the  mechanisms  that regulate  star
formation in low mass halos.


\acknowledgments  We thank  the anonymous  referee for  helpful  comments that
improved the presentation of this paper. This work is supported by 973 Program
(No.   2007CB815402),   the  CAS  Knowledge  Innovation   Program  (Grant  No.
KJCX2-YW-T05) and grants from  NSFC (Nos.  10533030, 10821302, 10925314).  HJM
would  like to acknowledge  the support  of NSF  AST-0607535.  LL  would thank
Gerard Lemson and  Yu Wang for their detailed  instructions on data retrieving
and analysing.  The Millennium Simulation databases used in this paper and the
web application  providing online access to  them were constructed  as part of
the activities of the German Astrophysical Virtual Observatory.


\end{document}